\documentclass[conference]{IEEEtran}

\usepackage{mathrsfs}

\ifCLASSINFOpdf
 
\else
  
\fi

\usepackage{amsthm}
\usepackage{makeidx}
\usepackage{examples}
\usepackage[pdftex]{graphicx}  
\usepackage{amssymb}
\usepackage{graphicx}
\usepackage{algorithm}
\usepackage{epstopdf}
\usepackage{color}
\usepackage{url}
\graphicspath{{../eps/}{../ps/}}
\usepackage{psfrag}    
\usepackage{algorithm}
\usepackage{algorithmic}
\usepackage{amsmath}
\font\msbm=msbm10 at 10pt

\newcommand{\FF}{\mbox{\msbm F}}

\newtheorem{theorem}{Theorem}

\newtheorem{remark}[theorem]{Remark}
\newtheorem{example}[theorem]{Example}
\newtheorem{defenition}[theorem]{Definition}

\IEEEoverridecommandlockouts
\begin{document}

\title{Reconstruction and Repair Degree of Fractional Repetition Codes\thanks{A one page abstract of this paper appears as a poster in IEEE Netcod 2013.}
}

\author{
\IEEEauthorblockN{Krishna Gopal Benerjee}
\IEEEauthorblockA{DA-IICT Gandhinagar\\
Email: krishna\textunderscore gopal@daiict.ac.in}
\and \IEEEauthorblockN{Manish K. Gupta}
\IEEEauthorblockA{DA-IICT Gandhinagar\\
Email: mankg@computer.org}
\and
\IEEEauthorblockN{Nikhil Agrawal}
\IEEEauthorblockA{DA-IICT Gandhinagar\\
Email: nikhil\textunderscore agrawal@daiict.ac.in}
}

\maketitle

\begin{abstract}
Given a Fractional Repetition (FR) code, finding the reconstruction  and repair degree in a Distributed Storage Systems (DSS) is an important problem. In this work, we present algorithms for computing the reconstruction and repair degree of FR Codes. 
\end{abstract}

\IEEEpeerreviewmaketitle
\section{Introduction}
Distributed Storage Systems (DSSs) use coding theory to provide reliability in the system. Recently a new class of regenerating codes known as "repair by transfer codes" were used to optimize disk I/O in the system \cite{shah2012distributed}. In this work, we consider DSS that use  Distributed Replication-based Simple Storage (DRESS) Codes consisting of an inner Fractional Repetition (FR) code and an outer Maximum Distance Separable (MDS) code to optimize various parameters of DSS \cite{rr10,Gupta:arXiv1302.3681}.  Codes which has rate at least the capacity of the system are known as universally good codes \cite{rr10}. To find out the universally good codes one has to find the reconstruction degree $k$ (minimum number of nodes one has to contact to yield the entire data) and the repair degree $d$ (number of nodes needs to be contacted in case of failure of a node)  in such a DSS. To the best of our knowledge there is no algorithm known for finding the reconstruction degree $k$ of a given FR code. It is easy to compute the repair degree $d$  for strong FR codes  as it is the degree of any node, however for weak FR codes no algorithm is known for computing the repair degree. Motivated by this in  this work, we present  algorithms for computing the reconstruction and repair degree of FR codes. 

This paper is organized as follows. Section $2$ collects necessary background on FR codes. In Section $3$ we present the algorithms for computing the reconstruction degree and in Section $4$ we present an algorithm for computing repair degree. 
Section $5$ concludes with general remarks.

\section{Background}
\begin{figure}
\centering
\includegraphics[scale=0.60]{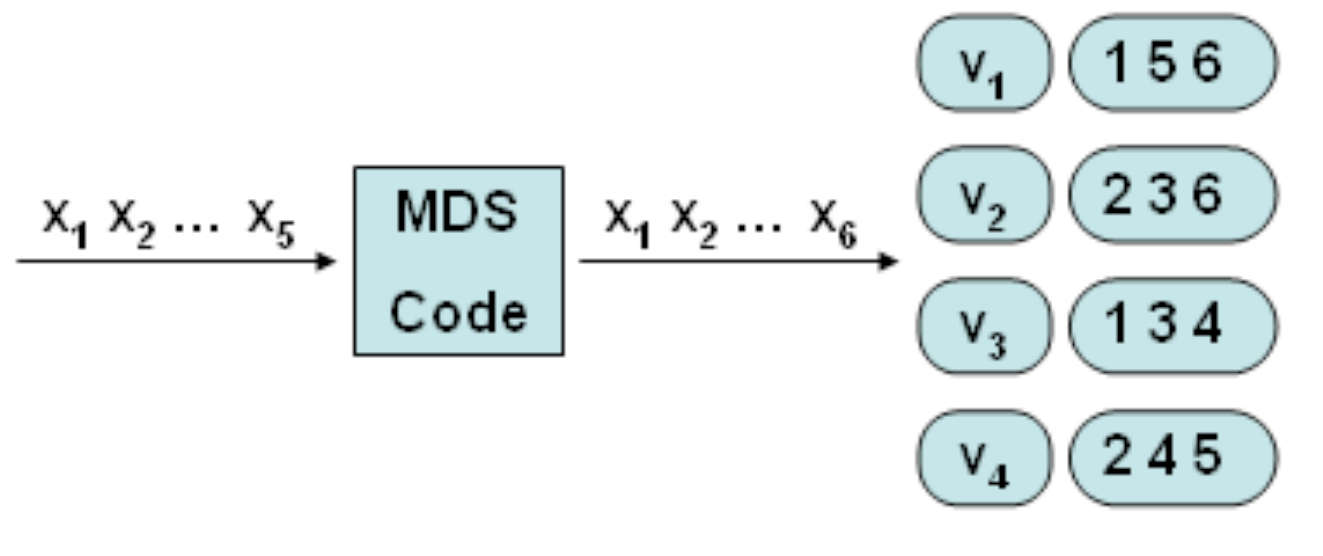}
\caption{DRESS Code consisting of fractional repetition code $\mathscr{C}$ having 4 nodes ($i.e.\ n=4$), 6 distinct packets ($i.e.\ \theta=6$),  repair degree $d=3$, replication factor $\rho=2$ and an outer MDS code}
\label{dress1}
\end{figure}

In an $(n,k,d)$ DSS, data is stored on $n$ nodes in such a fashion such that user can get the data by connecting any $k (k \leq n)$ nodes \cite{DBLP:journals/corr/abs-0803-0632}. In case of a failure of a node, data can be recovered by contacting any $d$ nodes and downloading few packets from them. This is achived by remarkable class of codes known as regenerating codes \cite{DARKYC11,DBLP:journals/corr/abs-0803-0632}, which optimizes the repair bandwidth as well as storage. However these codes fails to optimize disk I/O \cite{rr10}. Hence a new class of code known as "repair by tarnsfer codes" was introduced in \cite{shah2012distributed}.  This was further generalized to DRESS Codes consisting of an inner FR code and an outer MDS code to optimize various parameters of DSS \cite{rr10,Gupta:arXiv1302.3681}.  Figure  \ref{dress1} describes one such code where first a data file is divided into $5$ packes (usually elements of a finite field $\FF_q$), and then using a MDS code a parity packet is added. All these packets are now replicated two times on  $4$ nodes (such a replication is known as FR code) in such a way such that user can get entire data by contacting any $3$ nodes. Thus reconstruction degree is $3$. On the other hand if a node fails it can be repaired by contacting any $3$ nodes. Thus repair degree is $3$. Now we define formally FR codes and discuss some of its properties.

\subsection{Fractional Repetition codes}
FR code is an arrangement of  $\theta$ packets (each replicated $\rho$ times in a smart way) on $n$ nodes such that each node $U_i, 1 \leq i \leq n$ has $\alpha_i$ packets \cite{rr10,Gupta:arXiv1302.3681}.
\begin{defenition}
(Fractional Repetition Code): A Fractional Repetition (FR) code  denoted by $\mathscr{C}(n, \theta, \alpha, \rho)$ with replication factor $\rho$, for a DSS with parameter $(n,k,d)$, is a collection $\mathscr{C}$ of $n$ subsets $U_1,U_2, \ldots ,U_n$ of a set $\Omega = \{1,\ 2, \ldots, \theta \}$, which satisfies the following conditions:
\begin{itemize}
	\item Every member of $\Omega$ appears exactly $\rho$ times in the collection $\mathscr{C}$.
	\item $|U_i|=\alpha_i\ (\forall i = 1, 2, \ldots, n$ )
\end{itemize}
where $\alpha = \max \left\{ \alpha_i \right\} _{i=1}^n$.
\end{defenition}

Clearly, FR codes satisfy the equation (\ref{FRC}) \cite{Gupta:arXiv1302.3681}.

\begin{IEEEeqnarray}{rCl}
n \alpha=\rho  \theta + \delta,
\label{FRC}
\end{IEEEeqnarray}
where $ \theta$ packets are replicated $\rho$ times among $n$ nodes (each having weakness $\delta_i$) and $ \delta$ is total weakness of FR codes \cite{Gupta:arXiv1302.3681}. Thus $ \delta $ is given by $ \delta = \sum _{i=1}^n  \delta _i  = \sum _{i=1}^n \left( \alpha - \alpha_i \right)$ \cite{Gupta:arXiv1302.3681}.

\begin{remark}
For strong FR codes \cite{rr10}, $\delta = 0$ then equation  (\ref{FRC}) reduces to $n \alpha=\rho  \theta$, also in this case $\alpha_i = \alpha = d, \forall 1 \leq i \leq n$.
\end{remark}

\begin{example}
For FR code $\mathscr{C}(7, 8, 4, 3)$ a possible node packet distribution is shown in Table \ref{table:nonlin}.
\begin{table}[ht]
\caption{Node-Packet Distribution for FR code $\mathscr{C}:\ (7, 8, 4, 3)$}
\centering 
\begin{tabular}{|c|c|c|c|c|}
\hline
Nodes&Packets distribution&$\alpha_i$&$\delta_i=\alpha-\alpha_i$&$d_i$\\[0.5ex]
\hline\hline
$U_1$& 1, 6, 7, 8 &4&0&3\\ 
\hline
$U_2$& 1, 2, 7, 8 &4&0&2\\
\hline 
$U_3$& 1, 2, 3, 8 &4&0&3\\
\hline 
$U_4$& 2, 3, 4, 7 &4&0&3\\
\hline 
$U_5$& 3, 4, 5    &3&1&2\\
\hline
$U_6$& 4, 5, 6    &3&1&2\\
\hline
$U_7$& 5, 6       &2&2&1\\
\hline
\end{tabular}
\label{table:nonlin}
\end{table}
Note that in this example $\alpha =max\left\{4, 4, 4, 4, 3, 3, 2\right\}=4$, $\delta=\sum_{i=0}^7[\delta_i]$= 4 and it satisfies the relation $\ n \alpha = \rho \theta + \delta$.
\label{exmp}
\end{example}

\begin{defenition}
(Node-packet distribution incidence matrix): For FR code $\mathscr{C}(n, \theta, \alpha, \rho)$ a node-packet distribution incidence matrix \cite{DBLP:journals/corr/abs-1303-6801}  $M_{n\times\theta}$ is a matrix with its entries $m_{ij}$ given as 
\[
m_{ij}= \left\{
\begin{array}{cc}
0 & \mbox{if}\; j \in \Omega \;\mbox{s.t.}\; j \notin  U_i \\
1 & \mbox{if}\; j \in \Omega \;\mbox{s.t.}\; j \in U_i.
\end{array}
\right.
\]
The column support of each column $M_j, 1 \leq j \leq \theta$  of $M$ is denoted by $H_j= Supp (M_j)=\{ i | m_{ij} \neq 0\} $. We will need this in Section \ref{repair_di}.
\end{defenition}
\begin{remark}
Clearly, for a given FR code $\mathscr{C} \left( n, \theta, \alpha, \rho \right)$  its node-packet distribution incidence matrix $M_{n\times\theta}$ has the following properties. 
\begin{enumerate}
	\item Weight of $i^{th}$ row of  M is $\alpha _i$.
	\item Weight of each column of  M is $\rho$.
\end{enumerate}  
\end{remark}


\begin{example}
Node-packet distribution incidence matrix $M_{7\times8}$ for FR code $\mathscr{C}(7, 8, 4, 3)$  of Example \ref{exmp} is
\[
M_{7 \times 8} =
\begin{bmatrix}
1 & 0 & 0 & 0 & 0 & 1 & 1 & 1\\
1 & 1 & 0 & 0 & 0 & 0 & 1 & 1\\
1 & 1 & 1 & 0 & 0 & 0 & 0 & 1\\
0 & 1 & 1 & 1 & 0 & 0 & 1 & 0\\
0 & 0 & 1 & 1 & 1 & 0 & 0 & 0\\ 
0 & 0 & 0 & 1 & 1 & 1 & 0 & 0\\
0 & 0 & 0 & 0 & 1 & 1 & 0 & 0\\
\end{bmatrix}.
\]
\label{kg}
\end{example}
Given a $(n,k,d)$ DSS, one has to find a good FR code $\mathscr{C} \left( n, \theta, \alpha, \rho \right)$  which matches with the parameters of DSS. Note that the parameter $k$ in DSS is known as 
the reconstruction degree of DSS. If one wants to get the entire file one has to contact any $k$ nodes in DSS. However if we look at the definition of FR code $\mathscr{C}$ one finds that it is independent of 
$k$ (there is no direct formula for calculating the reconstruction degree). This motivates us to define the reconstruction degree of FR code $\mathscr{C}$ as the the number $k_{FR}$ so that if one wants the entire 
file (total $(\theta-1)$ packets as one remaining packet one can get using MDS code) one has to contact smallest set of any $k_{FR}$ nodes in FR code. Clearly, $k \leq  k_{FR}$. In order to find the value $k_{FR}$ of a FR code we also define another reconstruction degree $k^{\star}$ of FR code as the smallest subset of nodes of $\mathscr{C},$ that allows recovering the entire data (all $(\theta-1)$ packets). Clearly, we also have $k^{\star} \leq  k_{FR}$. We present an algorithm \ref{algofork} to compute $k^{\star}$. This gives a lower bound on actual $k_{FR}$. 

\begin{example}
Reconstruction degree $k^{\star}$ of the code $\mathscr{C}(7, 8, 4, 3)$ of Example \ref{exmp} is $2$ because using $U_2$ and $U_5$ one can get at least $7$ packets and $k_{FR}=4$ as contacting any $4$ nodes will give us at least $7$ packets. Another interesting example showing this difference is Figure $7$  of \cite{rr10}, where FR code $\mathscr{C}(6, 9, 3, 3)$ has $k^{\star}=3$ ($v_1, v_2$ and $v_3$ will give at least 8 packets) and $k_{FR}=4$ (any 4 nodes will give at least 8 packets). 
\end{example}
In Section \ref{algok}, we consider an algorithm for computing $k^{\star}$, and hence a lower bound on $k_{FR}$.  For a $(n,k,d)$ DSS one can define the rate of FR code \cite{rr10}.
\begin{defenition}
(Rate of FR Code): Given a $(n,k,d)$ DSS, rate $R_{\mathscr{C}}(k)$ of FR code $\mathscr{C}(n, \theta, \alpha, \rho)$ is defined as
\begin{equation*}
R_{\mathscr{C}}(k) =  \min { \bigcup _{i \in S}|U_i| }
\end{equation*}
$where \ S \subseteq \{1, 2,..., n \} \ and \ |S| = k$
\end{defenition}

It is clear that Rate $R_{\mathscr{C}}(k)$ of FR code $\mathscr{C}$ is the number of guaranteed distinct packets which an user will get when any $k$ nodes are contacted in $\mathscr{C}$. Thus finding 
the reconstruction degree $k$ is very useful in finding the rate of FR code. 

\begin{example}
Given a $(7,k,d)$ DSS, and a FR code $\mathscr{C}(7, 8, 4, 3)$ of Example \ref{exmp}, the rate of the code is $4$ for $k=3$ and rate is $6$ for $k=4$. 
\end{example}

\section{Algorithm for Computing Reconstruction degree $k$}\label{algok}

In order to find the reconstruction degree of a FR code, one can always delete one packet from all the nodes (W.L.O.G., we usually delete last packet $\theta$) as we can recover it using the parity of MDS codes.
Hence for constructing entire data  it is sufficient to reconstruct only $(\theta-1)$ packets. Thus WLOG we delete last packet $\theta$ in the algorithm \ref{algofork}.
\begin{algorithm}
\caption{Algorithm to compute reconstruction degree $k^{\star}$}
\begin{algorithmic}
\REQUIRE Node packet distribution of FR code after removing the last packet $\theta$ from all $n$ nodes of $V^n= \{ V_1, V_2, ..., V_n\}$.
\ENSURE $k^{\star}_{upp} =$ Reconstruction degree
\STATE$1:$ For $1 \leq i, j,m  \leq n$, if $\exists \ V_i \ \& \ V_j \ s.t. \ V_j \subseteq V_i$ then delete all such $V_j$ for all possible nodes $V_i$ and list remaining collection of nodes  as $V^m= \{ V_{i_1}, V_{i_2}, ..., V_{i_m}\}, |V_{i_j}|=\alpha_{i_j} = \;\mbox{number of packets in node}\; V_{i_j}.$
\STATE$2:$ Let $V^l$ = $ \left\{ V_{i_j} \in V^m | 1\leq j \leq m \ \& \ | V_{i_j}| = \max \{ \alpha_{i_j}\} \right\}$.
\STATE$3:$ Pick an arbitary set $V_{i_j} \in V^l$, and call this set as $P$. Set the counter $k_\lambda=1, 1 \leq k_\lambda \leq m$ and $1 \leq \lambda \leq |V^l|=l$.
\STATE$4:$ If $\exists \ V_{i_{j'}} (1 \leq {j}' \leq m) \in V^m \ s.t. \ V_{i_{j'}} \bigcap P = \phi$ then go to step $5$ otherwise jump to step $6$.
\STATE$5:$ Pick $V_{i_{j''}} (1 \leq {j''} \leq m) \in V^m $ which has max cardinality among all $V_{i_{j''}}$ in $V^m$ with $V_{i_{j''}} \bigcap P = \phi$. Update $P \ = \ P \bigcup V_{i_{j''}}$, update counter $k_\lambda  = (k_\lambda +1)$ and go to step $4$.
\STATE$6:$ If $\exists \ V_{i_r} (1 \leq r \leq m) \in V^m \ s.t. \ V_{i_r} \not\subset P$ then go to step $7$ otherwise go to step $8$.
\STATE$7:$ Pick  $V_{i_{r'}} (1 \leq r' \leq m) \in V^m $ which has maximum $\left| V_{i_{r'}}  \backslash  P \right|$ among all $V_{i_{r'}} \in V^m$ having the condition $V_{i_{r'}} \not\subset P$ then update $P \ = \ P \bigcup V_{i_{r'}}$, update counter $k_\lambda  = (k_\lambda +1)$  and  go to step $6$.
\STATE$8:$ If $ 1 \leq \lambda < l$, then store $k_\lambda$ in $k_\lambda^{\prime}$ and   set $k_\lambda = k_{(\lambda +1)}$ and 
perform step $4$ for   $P=V_{i_{j'''}} (1 \leq {j'''} \leq m) \in V^l  \ s.t.   V_{i_{j'''}} \neq V_{i_j} \in V^l$, otherwise report 
$k^{\star}_{upp} = \min\left\{ k_\lambda^{\prime} \right\}_{\lambda=1}^l$.
\end{algorithmic}
\label{algofork}
\end{algorithm}

We now consider an example to compute reconstruction degree $k^{\star}$ using algorithm \ref{algofork}.
\begin{example} 
Consider a FR code $(5, 9, 4, 2)$ as shown in Table \ref{exmpalgo1}.
\begin{table}[ht]
\caption{Node-Packet Distribution for FR code $(5, 9, 4, 2)$}
\centering 
\begin{tabular}{|c|c|}
\hline
Nodes&Packets distribution\\[0.5ex]
\hline\hline
$U_1$& 1, 2, 3, 4 \\ 
\hline
$U_2$& 1, 6, 9 \\
\hline 
$U_3$& 2, 5, 7, 9 \\
\hline 
$U_4$& 3, 5, 6, 8 \\
\hline 
$U_5$& 4, 7, 8  \\
\hline 
\end{tabular}
\label{exmpalgo1}
\end{table}

\begin{itemize}
	\item Note that  since $n=5$, after removing any packet (say last packet 9) we get $V^5 = \left\{ V_1, V_2, V_3, V_4, V_5 \right\}$, 
	where $V_1=\left\{ 1, 2, 3, 4 \right\}, \ V_2= \left\{ 1, 6 \right\}, \ V_3 = \left\{ 2, 5, 7 \right\}, \ V_4= \left\{3, 5, 6, 8 \right\}, \  V_5 = \left\{4, 7, 8 \right\}$ each having cardinality as $\{4,3,3,4,3\}$ respectively. 
	\item Further since there is no set $V_p \ s.t. \ V_p \subseteq V_q \\ (1 \leq p, q \leq 5)$ so step 1 yields \\ $V^m = V^5 = \left\{ V_1, V_2, V_3, V_4, V_5 \right\}$.
	\item For step 2, note that there are only two sets $V_1, V_4$ of maximum cardinality 4, so  $V^l =\left\{ V_1,  V_4 \right\}$ now executing step 3, 
	pick an arbitrary node $V_1$ as $P\ =\ V_1$ and initialize $k_1$ = 1.
	\item Now we skip step 4, since there does not exist any set $V_i \in V^5$ s.t. $V_i \bigcap P = \phi$ and we go to step 6.
	\item At step 6, we  search $V_i \in V^5$ s.t. $V_i \not\subset P$ so we get $V_1, V_2, V_3, V_4, V_5$.
	\item  For step 7 we have $V_2 \backslash P = \left\{ 6 \right\}$, $V_3 \backslash P = \left\{ 5, 7 \right\}$, $V_4 \backslash P = \left\{5, 6, 8 \right\}$ and $V_5 \backslash P = \left\{ 7, 8 \right\}$ among them $\left| V_4 \backslash P \right|$ is maximum. So $P$ = $P\bigcup V_4$ = $\left\{1, 2, 3, 4\right\}\bigcup\left\{3, 5, 6, 8 \right\}$\newline = $\left\{1, 2, 3, 4, 5, 6, 8\right\}$ and $k_1$ = 1+1 = 2.
	\item According to step 6, again we  search $V_i \in V^5$ s.t. $V_i \not\subset P$ and we get $V_3, V_5$. Now again $V_3 \backslash P = \left\{ 7 \right\}$ and $V_5 \backslash P = \left\{ 7 \right\}$.
	\item By step 7, P = $P\bigcup V_5$ = $\left\{1, 2, 3, 4, 5, 6\right\}\bigcup\left\{4, 7, 8\right\}$ = $\left\{1, 2, 3, 4, 5, 6, 7, 8\right\}$ and $k_1$ = 2+1 = 3 since $V_3 \backslash$ P is maximum.
	\item According to step 8, $k_1^{\prime}$  = 3  and update $k_1 = k_2 = 1$ Compute $k_2^{\prime}$ for $P$ = $V_4 \in V^2$, $k_2^{\prime}=3$.
	\item So $k^{\star}_{upp}$ = $\min \left\{ k_1^{\prime}, k_2^{\prime} \right\} = 3$
		\end{itemize}
\label{exm5942}
\end{example}

\begin{remark}
Note that in general, algorithm \ref{algofork} computes an upper bound on $k^{\star}$. However in Example  \ref{exm5942}, algorithm  gives an exact value of $k^{\star}$, i.e., $k^{\star}_{upp}= k^{\star}=3$.  Table \ref{ab} present a case of FR code $\mathscr{C}:\ (5, 8, 4, 2)$ for which $k^{\star}=2$ and $k^{\star}_{upp} =3$. Further note that at the cost of complexity, one can modify the algorithm \ref{algofork} at step 3, by taking $P$ on all possible nodes in $V_m$ to yield an exact reconstruction degree $k^{\star}$. In particular, for strong FR code this algorithm will always give an exact value of $k^{\star}$.

\end{remark}

\begin{table}[ht]
\caption{Node-Packet Distribution for FR code $\mathscr{C}:\ (5, 8, 4, 2)$}
\centering 
\begin{tabular}{|c|c|}
\hline
Nodes&Packets distribution\\[0.5ex]
\hline\hline
$U_1$& 1,2,3,4 \\ 
\hline
$U_2$& 1,2,5,7 \\
\hline 
$U_3$& 3,4,6,8 \\
\hline 
$U_4$& 7,8 \\
\hline 
$U_5$& 6 \\
\hline
\end{tabular}
\label{ab}
\end{table}
Arguments similar to Algorithm \ref{algofork}, can be used to give an algorithm for computing the exact reconstruction degree $k_{FR}$ as shown in Algorithm \ref{algoforkfr}. 

\begin{algorithm}
\caption{Algorithm to compute reconstruction degree $k_{FR}$}
\begin{algorithmic}
\REQUIRE A set of packets $\Omega =\{1,2, \ldots, \theta\}$ and node packet distribution of FR code with $n$ nodes $U^n= \{ U_1, U_2, \ldots, U_n\}$.
\ENSURE  Exact reconstruction degree $k_{FR}$.
\STATE$1:$ For  $1\leq m \leq n$ set $U^m= \{ U_1, U_2, ..., U_m\}$.  Take $m = n$. 
\STATE$2:$ Pick the set $U_m \in U^m$ and call this set as $P$. Set the counter $k_\lambda=1, 1 \leq k_\lambda \leq m$ and $1 \leq \lambda \leq n$. If $\Omega \backslash P = \phi$ or singleton set then go to step $6$ otherwise go to step $3$.
\STATE$3:$ If $\exists \ U_j (1 \leq j \leq m) \in U^m \ s.t. \ U_j \bigcap P = \phi$ then go to step $4$ otherwise jump to step $5$.
\STATE$4:$ Pick an arbitrary $U_{j'} (1 \leq {j'} \leq m) \in U^m $ which has maximum cardinality among all $U_{j'}$ in $U^m$ with $U_{j'} \bigcap P = \phi$. Update $P \ = \ P \bigcup U_{j'}$, update counter $k_\lambda  = (k_\lambda +1)$. Again if $\Omega \backslash P = \phi$ or singleton set then go to step $6$ otherwise go to step $3$.
\STATE$5:$ Pick  $U_{r} (1 \leq r \leq m) \in U^m \ s.t. \ U_r \not\subset P$ which has maximum $\left| U_r \backslash  P \right|$ among all $U_r \in U^m$ having the condition $U_r \not\subset P$ then update $P \ = \ P \bigcup U_{r}$, update counter $k_\lambda  = (k_\lambda +1)$. Once again if $\Omega \backslash P = \phi$ or singleton set then go to step $6$ otherwise go to step $5$.
\STATE$6:$ Stor $k_\lambda$ in $k_\lambda '$ and set $k_\lambda = k_{(\lambda +1)}$.
\STATE$7:$ If $1 \leq \lambda < n$ then calculate $U^{m-1}$ = $U^m \backslash \{ U_m \}$ and perform step $2$ for $P=U_{j''} (1 \leq {j''} \leq n) \in U^m$, otherwise report $k_{FR} = \max\left\{ k_\lambda ' \right\}_{\lambda=1}^n$.
\end{algorithmic}
\label{algoforkfr}
\end{algorithm}

In Section  \ref{repair_di}, we focus our attention to repair degree which is another important parameter of DSS.
\section{Algorithm for Computing Repair degree}\label{repair_di}

Given a $(n,k,d)$ DSS, in case of a node failure, it can be repaired by contacting any $d$ nodes \cite{rr10,DARKYC11}. Thus $d$ is known as the repair degree of a node. In case of FR codes, the repair of a node is Table based, i.e., one has to contact specific set of nodes for repair. However, in case of strong FR code $\mathscr{C}(n, \theta, \alpha, \rho)$, we have $\alpha = d$ for every node so it is easy to calculate the repair degree. Moreover, in case of weak FR code, if repair degree of a node $U_i, 1 \leq i \leq n$ is denoted by $d_i$ then $d_i \leq \alpha_i = |U_i| \leq \alpha$ since in the worst case all $\alpha_i$ packets can be recovered by contacting  some $\alpha_i$ nodes and $\alpha$ is maximum size of any node. As expected we also have $d_i \leq (n-1)$. A list of repair degree for all 5 nodes for FR code  $\mathscr{C}(7, 8, 4, 3)$  of Example \ref{exmp} is given in Table \ref{table:nonlin}. Note that the repair degree is much less than the number of packets in a node for weak FR code as compare to strong FR code where it is equal to the size of each node. Thus computing the repair degree of weak FR codes is an interesting problem. Algorithm \ref{algorepair} computes the repair degree $d_i$ for any node $U_i$. 


\begin{algorithm}
\caption{Algorithm to compute Repair Degree $d_i$}
\begin{algorithmic}
\REQUIRE Incidence matrix $M_{n \times \theta}$ of FR code.
\ENSURE Repair degree $d_i$ of node $U_i$.
\STATE$1:$ For each node $i, 1 \leq i \leq n$ let $S^{\{ i \}}_i = \{ H_j \backslash \{i\} | i \in H_j, 1 \leq j \leq \theta \}$
\STATE$2:$ Compute $T \subseteq \{ 1,2, \ldots, \theta \} \ s.t. \; |T|$ is maximum among all possible subsets and for $t \in T$,  $H_t \backslash \{i\} \in S^{\{ i \}}_i,$ and $ \bigcap H_t \backslash \{i\} \neq \phi$. 
Set counter $l_q (1 \leq q \leq n) = |T|-1.$
\STATE$3:$ Update $S^{\{ i \}}_i = S^{\{ i \}}_i \backslash (H_t \backslash \{i\}), \forall t \in T$.
\STATE$4:$ If $S^{\{ i \}}_i  = \phi$ then $d_i= \alpha_i -\sum_{\lambda=1}^q l_{\lambda},$ where $\alpha_i = |V_i|$, otherwise set $q=q+1$ and go to step $2$.
\end{algorithmic}
\label{algorepair}
\end{algorithm}

\begin{example} 
Consider the following node-packet distribution incidence matrix $M_{11 \times 8}$ for FR code $\mathscr{C}:\ (11, 8, 2, 3)$.
\[
M_{11 \times 8} =
\begin{bmatrix}
1 & 0 & 0 & 1 & 0 & 0 & 1 & 0\\
0 & 1 & 0 & 0 & 1 & 0 & 0 & 1\\
0 & 0 & 1 & 0 & 0 & 0 & 0 & 0\\
0 & 0 & 0 & 0 & 0 & 1 & 0 & 0\\
1 & 1 & 1 & 1 & 0 & 0 & 0 & 0\\ 
0 & 0 & 0 & 0 & 1 & 0 & 0 & 1\\
0 & 0 & 0 & 0 & 0 & 1 & 1 & 0\\
1 & 0 & 0 & 1 & 1 & 0 & 0 & 0\\
0 & 1 & 1 & 0 & 0 & 1 & 0 & 0\\ 
0 & 0 & 0 & 0 & 0 & 0 & 1 & 0\\
0 & 0 & 0 & 0 & 0 & 0 & 0 & 1\\
\end{bmatrix}.
\]

According to algorithm \ref{algorepair} the calculation of repair degree $d_i \ where \ i \in \{ 1, 2,...,11 \}$ for the node-packet distribution incidence matrix $M_{11 \times 8}$ is as follows.
\begin{itemize}
	\item $H_1$ = $\left\{1,8,5\right\}$, $H_2$ = $\left\{2,5,9 \right\}$, $H_3$ = $\left\{5,9,3\right\}$, $H_4$ = $\left\{8,1,5\right\}$, $H_5$ = $\left\{2,6,8\right\}$, $H_6$ = $\left\{9,4,7\right\}$, $H_7$ = $\left\{10,7,1\right\}$, $H_8$ = $\left\{2,6,11\right\}$.
	\item If we want to compute repair degree for $5^{th}$ node ($i.e. \ d_5$) then pick the all $H_j$ s.t. $5 \in H_j$ i.e. $H_1,\ H_2,\ H_3\ and\ H_4$.
	\item Now $ S^{\{ 5 \}}_5$ = $ \{H_1 \backslash \left\{5\right\}, H_2 \backslash \left\{5\right\}, H_3 \backslash \left\{5\right\}, H_4 \backslash \left\{5\right\} \}$,    where $H_1 \backslash \left\{5\right\}$ = $\left\{1,8\right\}$, $H_2 \backslash \left\{5\right\}$ = $\left\{2,9\right\}$, $H_3 \backslash \left\{5\right\}$ = $\left\{9,3\right\}$ and $H_4 \backslash \left\{5\right\}$ = $\left\{8,1\right\}$.
	\item But $\bigcap _{r \in \{1,2,3,4 \}} H_r \backslash \left\{5\right\} = \phi $ and there is no any common element among any three  sets chosen from the $ S^{\{ 5 \}}_5$. 
	\item But for T = $\{1, 4 \}$ we have $H_1 \backslash \left\{5\right\} \bigcap H_4 \backslash \left\{5\right\} = \left\{1,8 \right\} \neq \phi$ so $l_1$ = 2 - 1 = 1.
	\item Now updated $S^{\{ 5 \}}_5$ is $S^{\{ 5 \}}_5$ = $ \{H_2 \backslash \left\{5\right\}, H_3 \backslash \left\{5\right\} \}$then we have $H_2 \backslash \left\{5\right\} \bigcap H_3 \backslash \left\{5\right\} = \left\{9 \right\} \neq \phi$ here T = $\{2, 3\}$ so $l_2$ = 2 - 1 = 1.
	\item Now Repair degree ($d_5$) = $\alpha _5 - l_1 - l_2 = 2 \ where \ \alpha _5$ is weight of $5^{th}$ row in node-packet distribution incidence matrix $M_{11 \times 8}$.
\end{itemize}
\end{example}

\section{Conclusion}
In this paper, we presented algorithms for computing reconstruction degree of FR code $\mathscr{C}(n, \theta, \alpha, \rho)$. Given a FR code we define the reconstruction degree $k^{\star}$ as the smallest subset of nodes when contacted will give the entire data and provided algorithm for computing it. This gives a lower bound on the actual reconstruction degree $k_{FR}$ of FR code, which is defined as 
the, smallest number of any $k_{FR}$ nodes  when contacted will yield the entire data. At the cost of complexity, we also provided an algorithm for computing exact $k_{FR}$. Finally we show the significance of weak FR codes over strong FR codes using repair degree of FR codes. We also present an algorithm for computing repair degree for weak FR codes. 

\bibliographystyle{IEEEtran}
\bibliography{cloud}

\end{document}